\documentclass[final,5p,times,twocolumn,authoryear]{elsarticle}

\usepackage{amssymb}
\usepackage{xcolor}
\usepackage[colorlinks=true,linkcolor=black,citecolor=black,urlcolor=blue]{hyperref}

\newif\ifmarkchanges
\markchangesfalse
\ifmarkchanges
  \newcommand{\rev}[1]{{\color{blue}#1}}
\else
  \newcommand{\rev}[1]{#1}
\fi

\newcommand{\kms}{km\,s$^{-1}$}

\newcommand{\hmxbN}{53}
\newcommand{\hmxbslope}{1.02}
\newcommand{\hmxbslopeerr}{0.11}
\newcommand{\hmxbicpt}{-10.92}
\newcommand{\hmxbrms}{0.59}
\newcommand{\hmxbrho}{0.77}
\newcommand{\sigN}{23}
\newcommand{\sigslope}{1.16}
\newcommand{\sigslopeerr}{0.12}
\newcommand{\rppN}{152}
\newcommand{\rppslope}{0.72}
\newcommand{\rppslopeerr}{0.08}
\newcommand{\rpprms}{0.44}
\newcommand{\rpprho}{0.58}
\newcommand{\rppAN}{22}
\newcommand{\rppAslope}{1.10}
\newcommand{\rppAslopeerr}{0.16}

\newcommand{\rppCslope}{0.37}
\newcommand{\rppCslopeerr}{0.06}

\newcommand{\rppallN}{2351}
\newcommand{\rppallslope}{0.14}
\newcommand{\rppallslopeerr}{0.05}
\newcommand{\rppallrho}{0.04}
\newcommand{\atnfversion}{2.8.1}
\newcommand{\fittablerows}{%
SMC HMXB pulsars, all & 53 & $1.02\pm0.11$ & $-10.92\pm0.23$ & 0.59 & 0.77 \\
SMC HMXB pulsars, U+D only & 23 & $1.16\pm0.12$ & $-10.79\pm0.23$ & 0.47 & 0.91 \\
\hline
Young RPPs, $\tau_c<10^4$\,yr & 22 & $1.10\pm0.16$ & $-11.28\pm0.17$ & 0.31 & 0.83 \\
Young RPPs, $\tau_c<10^5$\,yr (adopted) & 152 & $0.72\pm0.08$ & $-12.44\pm0.07$ & 0.44 & 0.58 \\
Young RPPs, $\tau_c<10^6$\,yr & 528 & $0.37\pm0.06$ & $-13.34\pm0.04$ & 0.56 & 0.26 \\
All non-recycled RPPs & 2351 & $0.14\pm0.05$ & $-14.73\pm0.02$ & 1.04 & 0.04 \\}

\journal{Journal of High Energy Astrophysics}

\begin{document}

\begin{frontmatter}

\title{A possible evolutionary connection between high-mass X-ray binary pulsars and isolated rotation-powered pulsars}

\author{Jun Yang}
\ead{junyangpro@gmail.com}
\affiliation{organization={Department of Astronomy, Harvard University},
            addressline={60 Garden Street},
            city={Cambridge},
            state={MA},
            postcode={02138},
            country={USA}}

\begin{abstract}
We investigate the long-term spin evolution of high-mass X-ray binary (HMXB) pulsars in the Small Magellanic Cloud (SMC)
using archival measurements of spin periods and period derivatives from \textit{XMM-Newton}, \textit{Chandra}, and \textit{RXTE}. In the $P$--$|\dot P|$ plane,
the HMXB pulsars follow an approximately linear relation in log--log space with slope \rev{$\hmxbslope\pm\hmxbslopeerr$}, \rev{similar to the trend shown by the youngest} isolated rotation-powered pulsars \rev{(RPPs)} \citep[e.g.,][]{Manchester2005,Lorimer2005}.
\rev{For comparison, young RPPs from the ATNF Pulsar Catalogue give slopes that depend strongly on the characteristic-age cut ($\rppAslope\pm\rppAslopeerr$ for $\tau_c<10^4$\,yr and $\rppslope\pm\rppslopeerr$ for $\tau_c<10^5$\,yr), whereas the full non-recycled population shows no significant correlation. The HMXB pulsars lie in the same range of $\tau_c$ ($\sim1$--$10$\,kyr) as the youngest RPPs, but at spin periods two to three orders of magnitude longer.}
Reference characteristic-age contours in the HMXB $P$--$|\dot P|$ plane are roughly parallel to this relation, consistent with a continuous
evolutionary sequence.

\rev{If the measured long-term spin-up rates were sustained, slowly rotating accreting pulsars would reach spin periods typical of isolated radio
pulsars within only $\sim10^{2}$--$10^{4}$\,yr; for example, SXP\,101 ($P\approx101$\,s, $|\dot P|\approx1.9\times10^{-9}$\,s\,s$^{-1}$) would reach $P\sim1$\,s after $\sim2\times10^{3}$\,yr. Because these timescales are far shorter than the lifetimes of the Be donors, the spin at the end of the accretion phase is more likely set by torque equilibrium than by uninterrupted spin-up.} We suggest that a subset of isolated neutron stars may descend from accretion-powered HMXB pulsars
following binary disruption\rev{, most plausibly at the supernova of the donor star} \citep[e.g.,][]{Bhattacharya1991,TaurisTakens1998}, providing
an observational link between accretion-powered and rotation-powered neutron-star populations.
\end{abstract}


\end{frontmatter}

\section{Introduction}
Neutron stars are observed as accretion-powered X-ray pulsars, rotation-powered radio pulsars, and magnetars. Although evolutionary connections among these classes have long been discussed \citep{Bhattacharya1991}, observational constraints on pathways from accreting high-mass X-ray binaries (HMXBs) to isolated pulsars remain limited.

The Small Magellanic Cloud (SMC) hosts one of the richest known populations of Be/X-ray binaries \citep{Coe2015,Yang2017}. Long-term monitoring with \textit{RXTE}, \textit{XMM-Newton}, and \textit{Chandra} has yielded spin periods and period derivatives for dozens of pulsars over baselines of $\sim15$ yr \citep{Yang2017}, enabling direct tests of secular angular-momentum evolution in accreting neutron stars\rev{; see also \citet{Klus2014} for an independent analysis of spin-period changes in SMC Be/X-ray binaries}.

In this Letter, we compare the empirical locus of SMC HMXB pulsars in the $P$--$|\dot P|$ plane (with the sign of $\dot P$ encoded separately) with the classical $P$--$\dot P$ distribution of isolated pulsars\rev{, and in particular with that of young rotation-powered pulsars (RPPs)}. Our key result is that the long-term SMC measurements follow an approximately linear trend with slope near unity in log space, motivating the hypothesis that some accretion-powered pulsars could evolve toward the spin periods of isolated rotation-powered pulsars after binary disruption.

\section{Data and Analysis}

\rev{\subsection{SMC HMXB pulsars}}
We adopt spin period derivative measurements from the long-term SMC pulsar archive compiled by \citet{Yang2017}. \rev{The sample consists of the $\hmxbN$ pulsars with a measured long-term $\dot P$ in Table~3 of \citet{Yang2017}}, observed with \textit{XMM-Newton}, \textit{Chandra}, and \textit{RXTE} and spanning $P\sim\rev{2}$\,s to \rev{$\sim1300$\,s}. Following \citet{Yang2017}, we treat the reported derivatives as long-term, observation-averaged measures of secular behavior over the available baselines\rev{, converting the published values from ${\rm s\,day^{-1}}$ to ${\rm s\,s^{-1}}$. \citet{Yang2017} classify each source as a significant long-term spin-up (U; $\dot P/\sigma_{\dot P}\le-1.5$) or spin-down (D; $\dot P/\sigma_{\dot P}\ge+1.5$) system, or as consistent with no secular change (C).}

Figure~\ref{fig:hmxb} shows $|\dot P|$ versus $P$\rev{; the sign of $\dot P$ (net spin-up versus spin-down) and the significance class are encoded by the symbols.} A \rev{least-squares} linear regression in log space \rev{over all $\hmxbN$ pulsars} yields

\begin{equation}
\log |\dot P| = (\rev{\hmxbslope \pm \hmxbslopeerr})\log P + C,
\label{eq:fit}
\end{equation}

with \rev{$C=\hmxbicpt$ ($\dot P$ in s\,s$^{-1}$, $P$ in s), an rms scatter of $\hmxbrms$\,dex about the relation, and a Spearman rank correlation coefficient $\rho=\hmxbrho$. Restricting the fit to the $\sigN$ pulsars with a significant secular trend (classes U and D) gives a slope of $\sigslope\pm\sigslopeerr$ (Table~\ref{tab:fits}). This check is important because, for sources in class C, $|\dot P|$ is comparable to its uncertainty, which itself tends to increase with $P$; the fact that the significant subsample gives a consistent (slightly steeper) slope indicates that the relation is not produced by measurement noise alone.}

Characteristic-age contours are overplotted using the standard definition

\begin{equation}
\tau_c \sim \frac{P}{2|\dot P|}
\end{equation}

\citep[e.g.,][]{Lorimer2005}. Interestingly, the age contours are approximately parallel to the observed fitting relation\rev{; the HMXB locus lies close to $\tau_c\sim1$\,kyr}.

\rev{We also show illustrative contours of constant mass accretion rate $\dot M$ onto the neutron star. They assume that matter is accreted at the corotation radius, $R_{\rm mag}\approx R_{\rm co}=(GMP^2/4\pi^2)^{1/3}$, which is the condition for spin equilibrium, and that the resulting spin-up torque is $N\approx\dot M(GMR_{\rm co})^{1/2}$. With $2\pi I|\dot P|/P^2=N$, this gives $|\dot P|\propto\dot M P^{7/3}$, independent of the magnetic moment, for canonical $I=10^{45}$\,g\,cm$^2$ and $M=1.4\,M_\odot$.} In this framework, $\dot M$ is the accretion rate at the magnetosphere (i.e., not the donor mass-loss rate). \rev{The contours shown in Figure~\ref{fig:hmxb} span $\dot M=10^{-12}$--$10^{-9}\,M_\odot\,{\rm yr^{-1}}$} and are meant only as a guide under these simple torque assumptions.

\begin{figure*}
\centering
\includegraphics[width=0.75\textwidth]{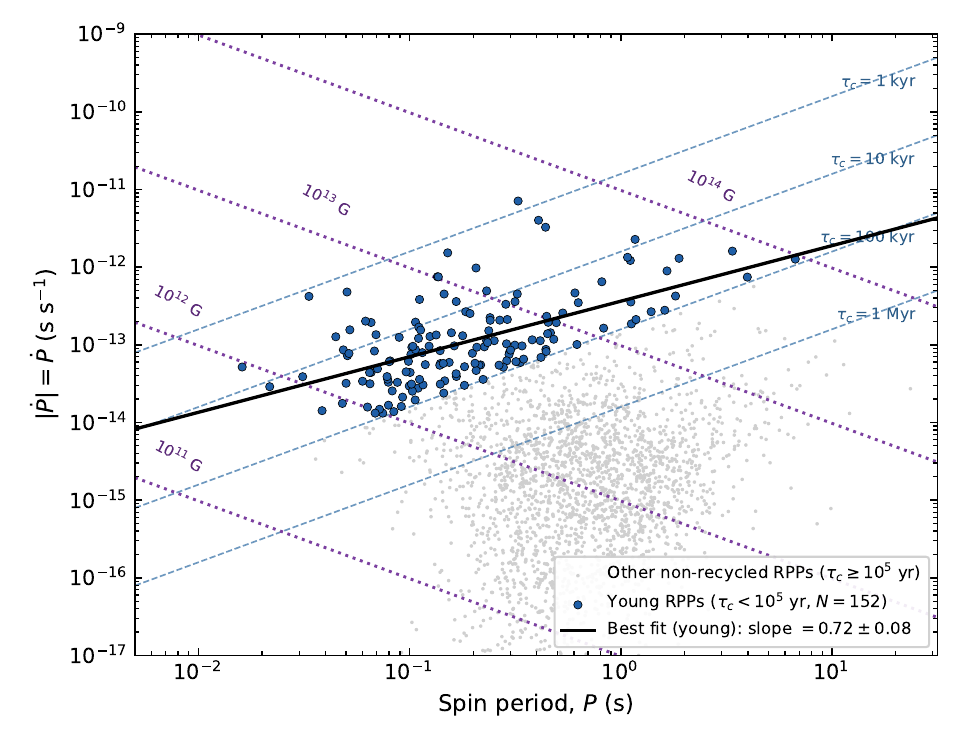}
\caption{\rev{$P$--$\dot P$ diagram of rotation-powered pulsars from the ATNF Pulsar Catalogue \citep{Manchester2005}, (version~\atnfversion), excluding magnetars, globular-cluster pulsars and recycled pulsars. Blue points are young RPPs ($\tau_c<10^5$\,yr, $N=\rppN$) and small gray points the remaining non-recycled pulsars. The black line is the best-fitting relation for the young RPPs (Table~\ref{tab:fits}). Blue dashed lines are characteristic-age contours and purple dotted lines magnetic-dipole field loci. For RPPs $\dot P>0$, so $|\dot P|=\dot P$.}}
\label{fig:rpp}
\end{figure*}

\rev{\subsection{Young rotation-powered pulsars}
For comparison, we took $P$ and $\dot P$ for all pulsars in the ATNF Pulsar Catalogue \citep{Manchester2005}. (version~\atnfversion). We excluded magnetars (sources classified as AXP or SGR), pulsars in globular clusters (whose $\dot P$ is contaminated by cluster accelerations) and recycled pulsars (dipole field $B<10^{10}$\,G), leaving $\rppallN$ non-recycled RPPs with $\dot P>0$. We define young RPPs as those with $\tau_c<10^{5}$\,yr ($N=\rppN$). Their $P$--$\dot P$ distribution is shown in Figure~\ref{fig:rpp}. The same log--log regression gives a slope of $\rppslope\pm\rppslopeerr$ with an rms scatter of $\rpprms$\,dex (Spearman $\rho=\rpprho$). Table~\ref{tab:fits} also lists the results for age cuts of $10^4$ and $10^6$\,yr and for the full non-recycled population. Any cut in $\tau_c$ is itself a line of unit slope in the $P$--$\dot P$ plane, so the fitted slope of an age-selected sample is partly shaped by the selection. This is why we report several cuts and the unselected population. Figure~\ref{fig:combined} places both populations and both best-fitting relations on identical logarithmic axes for direct comparison.}

\begin{table*}
\centering
\caption{\rev{Log--log linear fits $\log|\dot P| = a\log P + C$ ($\dot P$ in s\,s$^{-1}$, $P$ in s). $N$ is the number of sources, rms is the scatter about the fit in dex, and $\rho$ is the Spearman rank correlation coefficient. HMXB data: Table~3 of \citet{Yang2017}; RPP data: ATNF Pulsar Catalogue \citep[version~\atnfversion;][]{Manchester2005}, excluding magnetars, globular-cluster pulsars and recycled pulsars ($B<10^{10}$\,G).}}
\label{tab:fits}
\begin{tabular}{lccccc}
\hline
Sample & $N$ & Slope $a$ & $C$ & rms (dex) & $\rho$ \\
\hline
\fittablerows
\hline
\end{tabular}
\end{table*}

\section{Results}
The observed relation between $P$ and $|\dot P|$ in accreting pulsars resembles the classical $P$--$\dot P$ distribution of isolated rotation-powered pulsars \citep[e.g.,][]{Manchester2005,Lorimer2005}.
\rev{The quantitative comparison in Table~\ref{tab:fits} and Figure~\ref{fig:combined} shows how this resemblance should be qualified. For the youngest RPPs ($\tau_c<10^4$\,yr, $N=\rppAN$) the slope, $\rppAslope\pm\rppAslopeerr$, is consistent with the HMXB value of $\hmxbslope\pm\hmxbslopeerr$, and both populations occupy the same range of characteristic age: the HMXB locus lies close to $\tau_c\sim1$\,kyr, and the youngest RPPs have $\tau_c\sim1$--$10$\,kyr. However, the RPP slope decreases steadily as older pulsars are included ($\rppslope\pm\rppslopeerr$ for $\tau_c<10^5$\,yr and $\rppCslope\pm\rppCslopeerr$ for $\tau_c<10^6$\,yr). The full non-recycled population shows essentially no correlation (slope $\rppallslope\pm\rppallslopeerr$, Spearman $\rho=\rppallrho$). Because a cut in $\tau_c$ is a line of unit slope, a near-unity slope for the youngest RPPs is largely a consequence of the selection. The HMXB sample, by contrast, is defined by the detection of pulsations and a measured long-term $\dot P$ rather than by $\tau_c$, so its near-unity slope is an empirical property of the accreting population (see, however, the discussion of measurement noise in Section~2.1). The two populations are also offset in normalisation: at $P=1$\,s the HMXB relation lies $\sim1.5$\,dex above the relation for young RPPs with $\tau_c<10^5$\,yr, and the HMXB pulsars have spin periods two to three orders of magnitude longer than young RPPs of comparable $\tau_c$. The most robust statement is therefore that HMXB pulsars evolve along tracks of nearly constant effective spin-evolution timescale, $\tau_c\sim1$--$10$\,kyr, comparable to the characteristic ages of the youngest isolated pulsars.}

The similarity suggests that some HMXB pulsars may evolve into isolated pulsars after sustained long-term spin-up and subsequent binary disruption. One possible pathway is \rev{disruption of the binary at the supernova explosion of the donor star (Section~\ref{sec:disruption})} \citep[e.g.,][]{Bhattacharya1991,TaurisTakens1998}.

As an illustrative example, SXP\,101 currently has spin period

\begin{equation}
P \approx 101~{\rm s},
\end{equation}

and \rev{long-term} spin-up rate

\begin{equation}
\dot P \approx -1.6\times10^{-4}~{\rm s~\rev{day}^{-1}} \rev{\approx -1.9\times10^{-9}~{\rm s~s^{-1}}}.
\end{equation}

\rev{(We note that this rate is formally not significant, $\dot P/\sigma_{\dot P}=-0.48$; \citealt{Yang2017}.)} Assuming constant secular spin-up,

\begin{equation}
t \sim \frac{\Delta P}{|\dot P|},
\end{equation}

the pulsar could evolve to $P\sim1$ s within approximately

\begin{equation}
t \sim \rev{1.7\times10^3}~{\rm yr}.
\end{equation}

\rev{For the twelve pulsars with a significant long-term spin-up (class U), the same estimate gives $t\approx2\times10^{2}$--$5\times10^{3}$\,yr. These timescales are comparable to the characteristic ages of the youngest isolated pulsars, but they are very short compared with the remaining main-sequence lifetimes of the Be donors ($\gtrsim10^{6}$--$10^{7}$\,yr). Uninterrupted spin-up at the currently measured rates therefore cannot persist over the lifetime of the binary. As discussed below, the spin period at the end of the accretion phase is more plausibly set by the torque (quasi-)equilibrium reached for the late-time accretion rate.}

\section{Discussion}
\begin{figure*}
\centering
\includegraphics[width=0.75\textwidth]{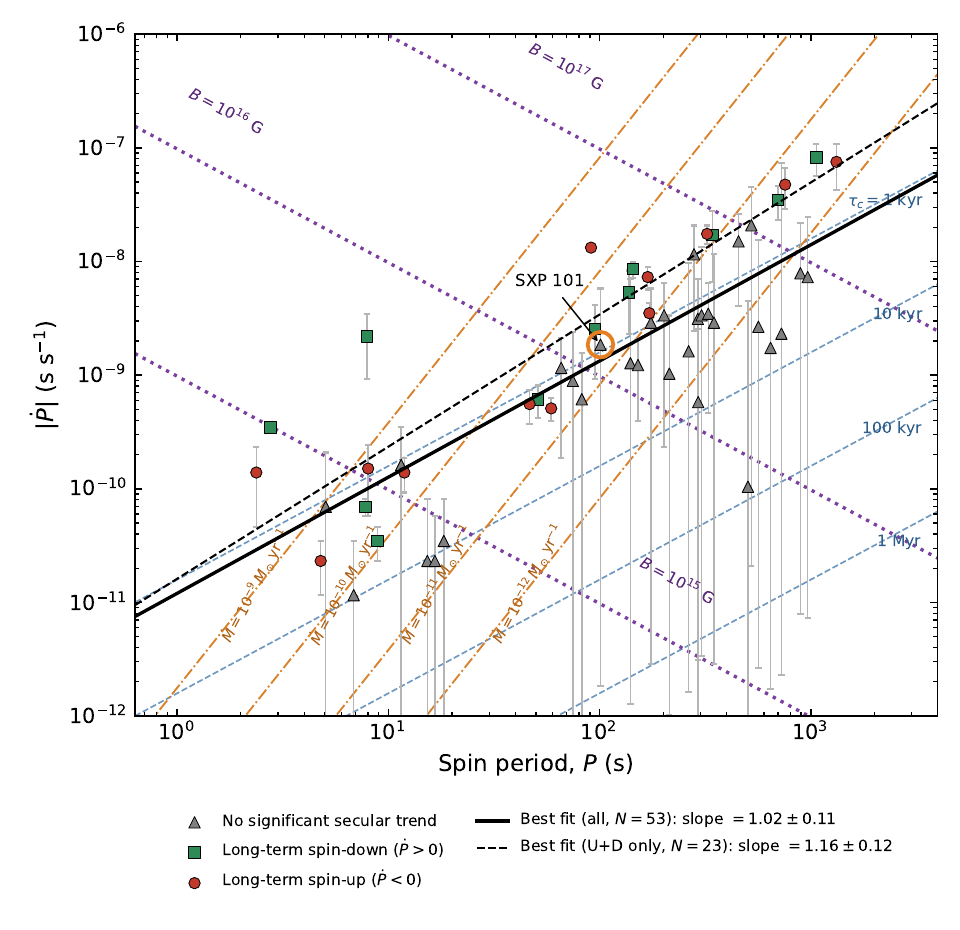}
\caption{\rev{SMC HMXB pulsars in the $P$--$|\dot P|$ plane (logarithmic axes; data from Table~3 of \citealt{Yang2017}). Red circles and green squares mark pulsars with a significant long-term spin-up ($\dot P<0$) and spin-down ($\dot P>0$), respectively; gray triangles mark sources without a significant secular trend. Gray bars show the $1\sigma$ uncertainties (lower bars truncated where they extend below zero). The solid black line is the best-fitting relation for all $\hmxbN$ pulsars (Eq.~\ref{eq:fit}) and the dashed black line the fit to the $\sigN$ significant (U+D) pulsars. Blue dashed lines are reference characteristic-age contours ($\tau_c=P/2|\dot P|$). Purple dotted lines are magnetic-dipole field loci ($B=3.2\times10^{19}(P|\dot P|)^{1/2}$\,G) from the isolated-pulsar framework, shown for comparison only. Orange dash-dotted lines are illustrative contours of constant accretion rate assuming $R_{\rm mag}=R_{\rm co}$ and $N\approx\dot M(GMR_{\rm co})^{1/2}$ with $I=10^{45}$\,g\,cm$^2$ and $M=1.4\,M_\odot$. The orange circle marks SXP\,101.}}
\label{fig:hmxb}
\end{figure*}

\begin{figure*}
\centering
\includegraphics[width=0.75\textwidth]{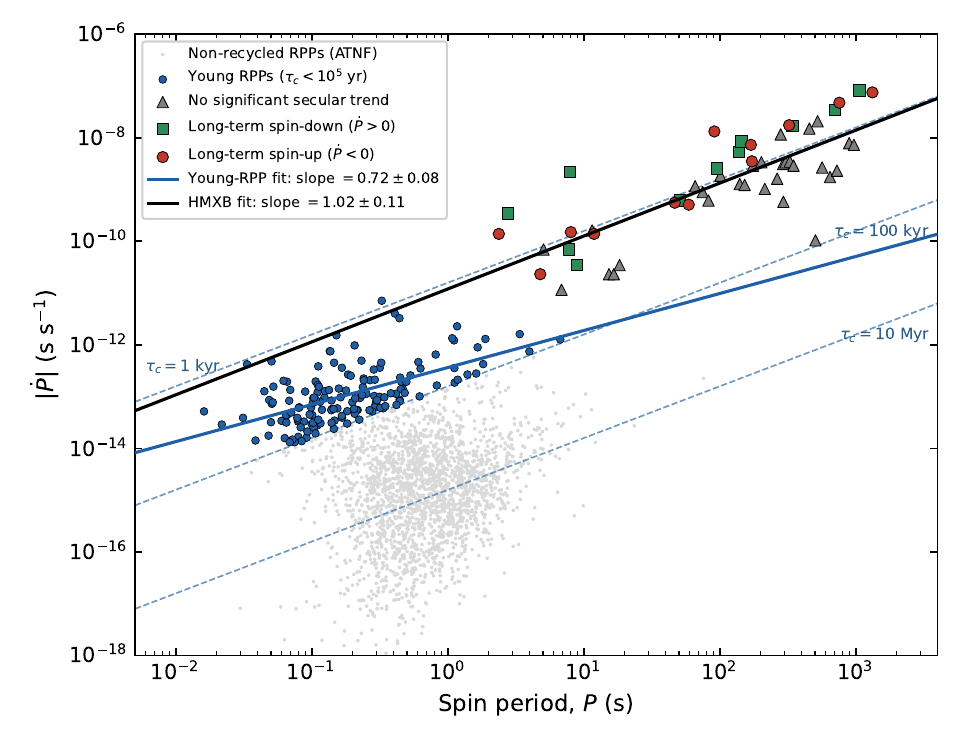}
\caption{\rev{SMC HMXB pulsars (symbols as in Fig.~\ref{fig:hmxb}) and rotation-powered pulsars (as in Fig.~\ref{fig:rpp}) on identical logarithmic axes. The black and blue lines are the best-fitting relations for the HMXB pulsars and the young RPPs, respectively (Table~\ref{tab:fits}). Blue dashed lines are characteristic-age contours.}}
\label{fig:combined}
\end{figure*}

Several caveats apply. First, accretion torques are unlikely to remain steady over Myr timescales: torque reversals, transient outbursts, and propeller phases can substantially modify the secular evolution \citep[e.g.,][]{Ghosh1979,Illarionov1975,Bildsten1997}\rev{. Indeed, 30 of the 53 SMC pulsars show no significant net $\dot P$ over $\sim15$\,yr \citep{Yang2017}, as expected for systems fluctuating about spin equilibrium.} Second, the efficiency and frequency of binary disruption (or donor detachment) in these systems are uncertain \rev{(Section~\ref{sec:disruption})}. Third, the measured $\dot P$ values reflect heterogeneous sampling and finite baselines, and the locus in Figure~\ref{fig:hmxb} should be interpreted as an empirical summary of long-term trends rather than an instantaneous torque relation.

A qualitative interpretation is possible within standard disk-accretion torque frameworks: if the long-term torque scales with a slowly varying combination of accretion rate and magnetic moment, then $\dot P$ can inherit an approximate power-law dependence on $P$ \citep[e.g.,][]{Ghosh1979}. The near-unity slope in the SMC sample therefore provides a compact empirical constraint for future torque modeling and population-synthesis calculations.

The characteristic-age contours in Figure~\ref{fig:hmxb} should be interpreted with care. For isolated pulsars, $\tau_c = P/(2\dot P)$ provides a useful spin-down timescale under magnetic-dipole braking assumptions \citep[e.g.,][]{Lorimer2005}. In HMXBs, however, $\dot P$ is primarily set by accretion torques rather than dipole radiation, so $\tau_c$ is not an estimator of the neutron-star age; instead, it can be viewed as an effective spin-evolution timescale for the observed torque state. For sources undergoing net spin-up, it is common to plot reference contours using $\tau_c \propto P/|\dot P|$ to avoid an unphysical negative timescale \citep[e.g.,][]{Ghosh1979,Bildsten1997}.

Similarly, magnetic-field loci commonly drawn on the isolated-pulsar $P$--$\dot P$ diagram (e.g., $B \approx 3.2\times10^{19}\sqrt{P\dot P}\,{\rm G}$) rely on the same dipole-braking framework \citep[e.g.,][]{Lorimer2005}. The magnetic-field contours shown in Figure~\ref{fig:hmxb} are therefore provided for visual comparison only and should not be interpreted as direct field measurements for the HMXB pulsars. \rev{Formally applying this relation to the HMXB pulsars would give $B\sim10^{15}$--$10^{17}$\,G, far above the $\sim10^{12}$--$10^{13}$\,G fields typically inferred from cyclotron lines in accreting pulsars, which shows that it cannot be used for these systems.} Field estimates for accreting pulsars require additional assumptions (e.g., torque models or cyclotron-line measurements).

The near-parallelism between characteristic-age contours and the empirical HMXB relation may nonetheless indicate that long-term accretion-torque evolution preserves a scaling similar to that produced by magnetic-dipole braking in isolated pulsars. Population-synthesis modeling will be useful for assessing whether the observed HMXB locus can plausibly evolve into the isolated pulsar distribution.

\rev{\subsection{Likelihood of binary disruption along the proposed channels}
\label{sec:disruption}
The scenario requires the accreting neutron star to become isolated (or at least to stop accreting). Below we assess the main channels qualitatively.

\textit{(i) Disruption at the supernova of the donor.} The SMC Be/X-ray binaries have already survived the first supernova, so the relevant event is the core collapse of the Be donor at the end of its life. For circular orbits, a binary becomes unbound if the instantaneous, symmetric mass loss exceeds half of the total pre-explosion mass \citep{Blaauw1961}. For a donor of $\sim8$--$20\,M_\odot$ that collapses to a $\sim1.4\,M_\odot$ remnant with little prior mass loss, this condition is easily met. In addition, natal kicks of typically a few hundred \kms\ \citep{Hobbs2005} are comparable to or larger than the relative orbital velocities of Be/X-ray binaries ($\sim80$--$250$\,\kms\ for $P_{\rm orb}\sim10$--$300$\,d and a total mass of $\sim16\,M_\odot$). Analytic and numerical studies of asymmetric supernovae in binaries \citep{Hills1983,Brandt1995,Kalogera1996,TaurisTakens1998} therefore imply that the large majority of such systems are unbound at the second supernova, and only a minority survive as double neutron stars. This channel releases the first-born (formerly accreting) neutron star as an isolated object, with a modest space velocity of the order of its former orbital velocity \citep{TaurisTakens1998}. We therefore regard it as the dominant and most probable route. Its main limitation is timing: it operates only after $\sim10^{6}$--$10^{7}$\,yr, long after the spin-up timescales estimated above. The spin of the released neutron star thus reflects the late-time torque equilibrium rather than extrapolated spin-up.

\textit{(ii) Donor detachment without disruption.} Accretion can also stop while the binary stays bound, for example when the Be decretion disk dissipates or the orbit is too wide for efficient wind capture. The neutron star is then not isolated, but it could be observable as a rotation-powered pulsar with a massive companion. Radio pulsars with B/Be companions are known, including PSR J0045$-$7319 in the SMC \citep{Kaspi1994} and PSR B1259$-$63 \citep{Johnston1992}. These systems are, however, usually interpreted as neutron stars that have not yet begun (or are not currently) accreting, and they illustrate the link between the two classes more directly in the opposite sense (RPP to HMXB). Detachment is probably frequent on short timescales (e.g., quiescent phases of Be/X-ray transients) but does not by itself produce an isolated pulsar.

\textit{(iii) Common-envelope evolution and mergers.} In the closest systems, expansion of the donor can lead to unstable mass transfer, a common-envelope phase and possibly a merger. These outcomes do not produce isolated rotation-powered pulsars and are expected to affect only a small fraction of Be/X-ray binaries, whose orbits are typically wide.

\textit{(iv) Dynamical interactions.} Encounters with other stars are negligible for field binaries in the SMC, given their low stellar densities.

Two further points limit how many HMXB pulsars could end up in the observed RPP population. First, most isolated neutron stars are probably produced by disruption at the \emph{first} supernova \citep[e.g., $\sim$86\% of massive binaries in the models of][]{Renzo2019}, and these never go through an accreting phase. HMXB descendants would therefore form a subset of isolated neutron stars, as proposed here, rather than the dominant channel. Second, a neutron star released with a spin period of tens to hundreds of seconds and a field of $\sim10^{12}$\,G lies beyond the radio pulsar death line \citep{ChenRuderman1993}. Only systems released at short spin periods, for example after a phase of high accretion rate, would join the observable population of radio pulsars. A quantitative estimate of these probabilities requires population synthesis, which is beyond the scope of this Letter.}

\section{Conclusions}
We find that SMC HMXB pulsars exhibit an approximately linear trend in the $P$--$|\dot P|$ plane \rev{(slope $\hmxbslope\pm\hmxbslopeerr$)} that resembles the classical $P$--$\dot P$ distribution of isolated rotation-powered pulsars \rev{(the youngest RPPs, $\tau_c<10^4$\,yr: slope $\rppAslope\pm\rppAslopeerr$; the slope for RPPs depends strongly on the age cut and vanishes for the full population)}, suggesting a potential evolutionary link between accretion-powered and rotation-powered neutron stars.

If net long-term spin-up \rev{at the measured rates} persisted, slowly rotating HMXB pulsars could reach $P\sim1$ s on timescales of \rev{only $\sim10^{2}$--$10^{4}$\,yr. Because these timescales are much shorter than the lifetimes of the donors, the spin at the end of accretion is more likely set by torque equilibrium. Disruption of the binary at the supernova of the donor, the most probable channel, could then yield isolated neutron stars.}

Future timing campaigns and population-synthesis studies will be essential for testing this scenario.

\section*{Acknowledgements}
We thank the anonymous referee for helpful comments\rev{ that improved the paper}. This work made use of archival observations from \textit{XMM-Newton}, \textit{Chandra}, and \textit{RXTE}\rev{, and of the ATNF Pulsar Catalogue \citep{Manchester2005}.The author gratefully acknowledges the generous philanthropic support}.

\section*{Data Availability}
The results in this paper are based on analysis of archival \textit{XMM-Newton}, \textit{Chandra}, and \textit{RXTE} observations. The \textit{XMM-Newton} data are available from the XMM-Newton Science Archive (XSA), the \textit{Chandra} data from the Chandra Data Archive (CDA), and the \textit{RXTE} data from the HEASARC archive. \rev{The HMXB spin-period derivatives used here are published in Table~3 of \citet{Yang2017}, and the rotation-powered pulsar parameters are publicly available from the ATNF Pulsar Catalogue (\url{https://www.atnf.csiro.au/research/pulsar/psrcat/}).} Additional analysis products and scripts will be made available by the corresponding author upon reasonable request.

\bibliographystyle{elsarticle-harv}
\bibliography{references}

\end{document}